\documentclass[12pt,a4paper]{article}
\usepackage{graphicx}
\usepackage{amsmath,amssymb}
\usepackage{authblk}

\newcommand{\tr}{{\rm Tr}}
\newcommand{\Wg}{{\rm Wg}}

\newtheorem{proposition}{Proposition}

\newcommand{\be}{\begin{equation}}
\newcommand{\ee}{\end{equation}}

\begin{document}

\title{Commutators of random matrices from the unitary and orthogonal groups}
\author{Pedro H.S. Palheta, Marcelo R. Barbosa, Marcel Novaes}
\affil{\normalsize Instituto de F\'isica, Universidade Federal de Uberl\^andia, 38408-100, Brazil}
\date{}

\maketitle
\begin{abstract}
We investigate the statistical properties of $C=uvu^{-1}v^{-1}$, when $u$ and $v$ are independent random matrices, uniformly distributed with respect to the Haar measure of the groups $U(N)$ and $O(N)$. An exact formula is derived for the average value of power sum symmetric functions of $C$, and also for products of the matrix elements of $C$, similar to Weingarten functions. The density of eigenvalues of $C$ is shown to become constant in the large-$N$ limit, and the first $N^{-1}$ correction is found. 

\end{abstract}

\section{Introduction}

The unitary and orthogonal groups, $U(N)$ and $O(N)$, are central to physics and mathematics in general. Because they have a unique normalized positive invariant measure, known as Haar measure, they can be seen as probability spaces and it is natural to enquire about various probability distributions associated with them. The joint distribution of eigenvalues, for example, was already known to Weyl \cite{weyl}. Symmetric polynomials in the eigenvalues have attracted a good deal of attention \cite{diaconis1,johanson,diaconis2}, as well as characteristic polynomials \cite{keating,conrey,freezing}, notably as models for the Riemann zeta function and other $L$-functions. A broad and accessible account can be found in \cite{meckes}.

The subject was brought into physics by Dyson in a series of papers in 1962 \cite{dyson}, although what he then called the `orthogonal ensemble' is not the orthogonal group. Another major source of interest was the introduction of the Itzykson-Zuber integral in 1980 \cite{IZ}, which was later realized to be a particular case of the Harish-Chandra integral \cite{HC}. Integrals of this kind continue to generate important investigations \cite{zinn,zeitouni,zuber,segala,guay}.

The distribution of the modulus squared of any matrix element is easily obtained from an invariance argument, but joint distributions of matrix elements are not yet available (it is known that small subblocks have a Gaussian distribution in the large $N$ limit). Apparently the first to consider averages of products of matrix elements was Weingarten \cite{wein}, and they were later used in a variety of contexts \cite{mello,brouwer,lam}. The fundamental quantities is this respect are nowadays called Weingarten functions \cite{collins,CS,CM,Banica1,Banica,matsumoto,zinn2}. Their large dimension expansions have been related to many different combinatorial problems related to factorizations of permutations \cite{MN,marcel,mat,BK}. 

In this work we explore the statistical properties of the group commutator, $C(u,v)=uvu^{-1}v^{-1}$, when $u$ and $v$ are independent random unitary or orthogonal matrices. If $u$ and $v$ commute, then $C(u,v)=1$. 

A different notion of commutator, $[A,B]=AB-BA$, is of course of fundamental importance in quantum mechanics, as the Heisenberg uncertainty principle $\Delta q \Delta p\ge \frac{\hbar}{2}$ is intimately connected with the canonical commutation relation $[q,p]=i\hbar$. This commutator is related to the group commutator because, if we take $u=e^{-tA}$ and $v=e^{-tB}$, then 
\be uvu^{-1}v^{-1}=e^{t^2[A,B]+O(t^3)}.\ee In the same vein, after the seminal work \cite{otoc1}, the so-called out-of-time-order correlators, which are related to commutators of the form $[A(t),B(0)]$ as a function of a time variable $t$, have been attracting increasing attention as important measures of quantum chaos or quantum complexity (see \cite{otoc2,otoc3,otoc4,otoc5,otoc6}, for example). Instead of treating these sophisticated correlators, we address simpler and more basic questions: how is the commutator distributed? What are its statistical properties?

Concretely, we define matrix probability spaces consisting of commutators inside the unitary and orthogonal groups, denoted
\be CG(N)=\{uvu^{-1}v^{-1},u,v\in G(N)\}.\ee The probability measure on $CG(N)$ is induced from the Haar measure of $G(N)$, with $G\in\{U,O\}$. Because a commutator is a complicated object, with many degrees of freedom, understanding its distribution completely is a challenge. But we can answer some questions. Our results about the statistical properties of the eigenvalues and matrix elements of commutators are presented in Section 2.

This topic has arisen before, although not so directly as we do here. An extensive study of average values of traces of `words' on Lie groups and the combinatorics of their large-$N$ expansions was carried out by Magee and Puder in \cite{magee1,magee2,magee3}. An important role in that analysis is what they call the commutator length of the words. In particular, the average trace of $C^n$ was computed in the unitary case up to $n=3$ and for the orthogonal case up to $n=2$. Products of independent random unitary and orthogonal matrices were also considered in \cite{radulescu,mingo1,mingo2,redelmeier}, in the context of second order freeness. In particular, the results from these works imply that $\tr (C)$ has a Gaussian distribution for large $N$, in both cases. We rederive this fact.

In the present contribution to this topic, we obtain results about the average value of symmetric functions of the eigenvalues of $C$ and for the average value of general polynomials in the matrix elements of $C$, within $CU(N)$ and $CO(N)$. From those results we extract the first two orders in $1/N$ of the density of eigenvalues for these ensembles. For simplicity, we avoid the treatment of unitary symplectic groups $Sp(N)$, but that analysis proceeds in analogy with the orthogonal one, and in fact results in that case may be recovered from the well known duality $Sp(N)\leftrightarrow O(-2N)$ \cite{magee3,veselov,mulase}.

We state our results in Section 2. Proofs and discussion are presented for the unitary group in Section 3 and for the orthogonal group in Section 4. Required facts about permutation groups, the Brauer monoid and universal characters of orthogonal groups, including all the notation and terminology used to state the results, are collected in two Appendices. In our numerical simulations we used the algorithm suggested by Mezzadri \cite{mezzadri}.

\section{Results}

We start by considering the average value of symmetric functions of the eigenvalues of $C$. In particular, let 
\be p_\mu(C)=\prod_{i=1}^{\ell(\mu)}\tr (C^{\mu_i})\ee be the power sum symmetric functions.
Then we have 

\begin{proposition} For $\mu\vdash n$, let $\chi_\lambda(\mu)$ be irreducible characters of the permutation group $S_n$ and $b_{\lambda}(\mu)$ be what we call universal Brauer characters. Then
\be \langle p_\mu(C)\rangle_{CU(N)}=n!\sum_{\substack{\lambda\vdash n\\\ell(\lambda)\le N}} \frac{\chi_\lambda(\mu)}{d_\lambda [N]_\lambda}\ee
and 
\be \langle p_\mu(C)\rangle_{CO(N)}=\sum_{h=0}^{\left \lfloor{n/2}\right \rfloor} (n-2h)!\sum_{\substack{\lambda\vdash n-2h\\\ell(\lambda)\le N}}\frac{b_{\lambda}(\mu)}{d_\lambda \{N\}_\lambda},\ee
where $d_\lambda=\chi_\lambda(1^n)$, while $[N]_\lambda$ and $\{N\}_\lambda$ are explicit polynomials, closely related to the dimensions of irreducible representations of the Lie groups, given respectively in Eq. (\ref{Nl}) and Eq. (\ref{Nol}).
\end{proposition}
In both cases, it is easy to see that $\langle \tr(C)\rangle=1/N$.

For the orthogonal group, $\langle p_{1^n}(C)\rangle$ can be interpreted as moments of the probability density of $\tr (C)$. As we discuss in the proof of Proposition 1, our results imply that for $CO(N)$ with large $N$, that probability density is asymptotically Gaussian and given by $\frac{1}{\sqrt{2\pi}}e^{-(x-\frac{1}{N})^2/2}$. This was already known \cite{mingo1,mingo2,redelmeier}.

For the unitary group, what follows from Proposition 1 is that $\langle (\tr(C))^n\rangle_{CU(N)}=\frac{n!p(n)}{N^n}+O(N^{-n-1}),$ where $p(n)$ is the number of partitions of $n$. However, in this case $\tr (C)$ is a complex variable whose distribution cannot be fully characterized by $\langle p_{1^n}(C)\rangle$. It is known that the asymptotic distribution is of a complex Gaussian \cite{mingo1}. 

The eigenvalue density of random unitary or orthogonal matrices is constant on the unit circle for any value of $N$. What about the eigenvalue density of their commutators? Eigenvalues are of course of the form $e^{i\theta}$ for $0\le \theta <2\pi$. We denote by $\rho_{CG(N)}(\theta)$ the probability density of the eigenphase for random matrices in $CG(N)$. When $N$ is odd, every matrix in $CO(N)$ has one eigenvalue equal to $1$; the density we consider is for the remaining ones. Since $C$ and $C^{-1}$ only differ by the exchange of the factors $u$ and $v$, they are equally likely, so $\rho_{CG(N)}(\theta)$ is an even function.

\begin{figure}[t!]
\includegraphics[scale=0.8,clip]{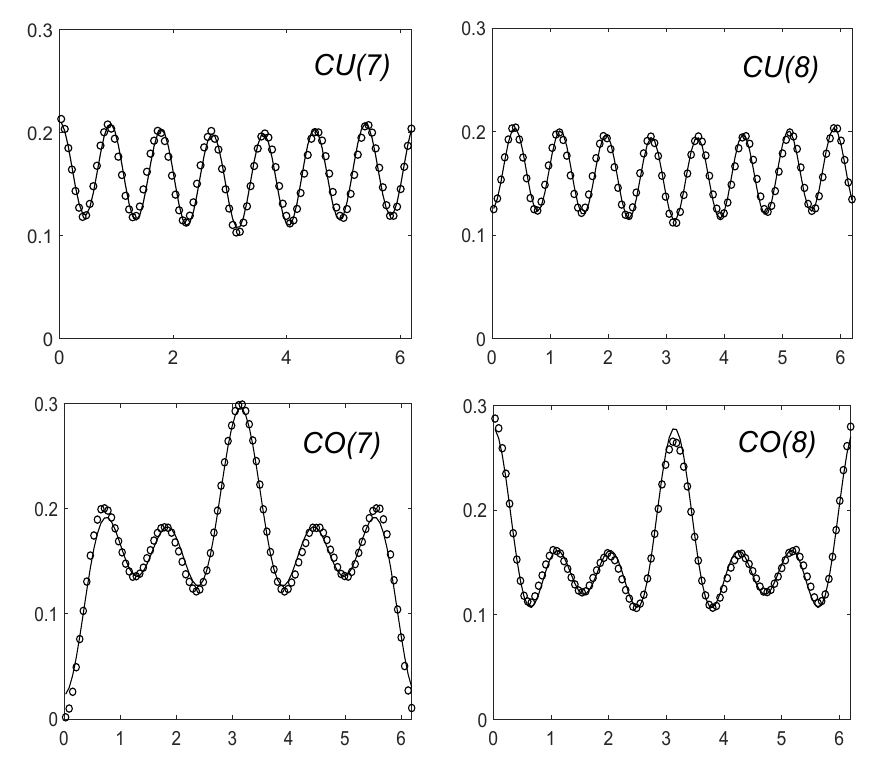}
\caption{Eigenphase density for the commutator in $CU(7)$ and $CU(8)$ (up) and in $CO(7)$ and $CO(8)$ (down). Average value is $(2\pi)^{-1}\approx 0.16$. Circles are numerical results (in each case, from a sample of $10^6$ random matrices), solid lines are theoretical asymptotics as given in Proposition 2.} 
\end{figure}

In contrast with $G(N)$, the eigenphase density for $CG(N)$ is not constant for finite $N$. In Figure 1 we present a few  numerically obtained eigenphase distributions of $CU(N)$ and $CO(N)$ (circles). In fact, we show

\begin{proposition}  The density of eigenphases of $C$ for $CU(N)$ is given by 
\be \rho_{CU(N)}(\theta)=\frac{1}{2\pi}+\sum_{n=1}^\infty c_{N,n}\cos(n\theta),\ee
 where 
\be c_{N,n}=\frac{n}{N^2} \sum_{k=0}^{{\rm min}(n,N)-1}(-1)^k{N-1 \choose k}^{-1}{N+n-k-1 \choose n-k-1}^{-1}.\ee
Its asymptotics for large $N$ is
\be\label{arho1}  \rho_{CU(N)}(\theta)=\frac{1}{2\pi}-\frac{(-1)^N}{N\pi}\cos(N\theta)+O(N^{-2}).\ee 
For the $CO(N)$, the asymptotic behavior is
\be \rho_{CO(N)}(\theta)=\frac{1}{2\pi}-\frac{1+(-1)^N}{4\pi N}+\frac{(-1)^N}{2N\pi}\frac{\sin((N-1)\theta)}{\sin(\theta)}+O(N^{-2}).\ee
\end{proposition}

This approximation is shown as solid lines in Figure 1. The leading $N^{-1}$ correction suggests, in the unitary case, that the eigenphases tend to cluster around $N$th-roots of unit if $N$ is odd and to avoid them if $N$ is even. In the orthogonal case with $N$ odd, repulsion by the fixed eigenvalue at $\theta=0$ leads to a depletion of the density at the origin and an increase at the antipode point.

Let us also remark that the eigenvalues of commutators repel each other, as one would in general expect for random matrices. We observed numerically (not shown) that the eigenphase spacing distribution for $CG(N)$ conforms well to the Wigner surmise, as is the case for $G(N)$. 

We now turn to the average value of polynomials in the matrix elements of $C$. For the unitary group, polynomial integrals may be expressed as linear combinations of terms like \cite{collins}
\be\label{weingU}\left\langle \prod_{t=1}^n u_{i_tj_t}u^*_{q_tp_t}\right\rangle_{U(N)} = \sum_{\sigma,\tau \in S_n} \delta_\tau(\vec{q},\vec{i})\delta_\sigma(\vec{p},\vec{j})\Wg^{U}_N(\sigma^{-1}\tau), \ee where $\delta_\sigma(\vec{i},\vec{j})=\prod_{k=1}^n \delta_{i_k,j_{\sigma(k)}}$ and $\Wg^{U}_N$ are the called the Weingarten functions.

For the commutator in the unitary group, we have

\begin{proposition}
Let $\vec{i}$ and $\vec{j}$ be two lists of $n$ numbers between $1$ and $N$. Then
\be\label{CU} \left\langle \prod_{k=1}^n C_{i_k,j_k}\right\rangle_{CU(N)}=\sum_{\pi\in S_n}\delta_\pi(\vec{i},\vec{j})F_N^U(\pi),\ee where \be\label{FU} F_N^U(\pi)=n!\sum_{\substack{\lambda\vdash n\\\ell(\lambda)\le N}}\frac{\chi_\lambda(\pi)}{d_\lambda [N]_\lambda^2}.\ee
\end{proposition}

For the orthogonal group, polynomials integrals can be written in terms of \cite{CS,CM}
\be\left\langle \prod_{t=1}^{2n} u_{i_tj_t}\right\rangle_{O(N)} = \sum_{\sigma,\tau \in M_n} \Delta_\tau(\vec{i})\Delta_\sigma(\vec{j})\Wg^{O}_N(\sigma^{-1}\tau), \ee  where the sum is over matchings (see Appendix A), the function $\Delta_\tau(\vec{i})$ equals $1$ if the string $\vec{i}$ satisfies the matching $\tau$, and vanishes otherwise (see Eq. (\ref{Delta})), and $\Wg^{O}_N(\sigma)$ are the corresponding Weingarten fuctions. 

For the commutator in the orthogonal group, we managed to obtain an explicit formula only when the lists $\vec{i}$ and $\vec{j}$ have no repeated indices. In that case, they must be related by a single permutation and we get

\begin{proposition} Let $\pi \in S_n$ and $\vec{i}$ a list with $n$ distinct entries. Then,
\be\left\langle \prod_{k=1}^n C_{i_k,i_{\pi(k)}}\right\rangle_{CO(N)}=F_N^O(\pi)=n!\sum_{\substack{\lambda\vdash n\\\ell(\lambda)\le N}}\frac{\chi_\lambda(\pi)}{d_\lambda\{N\}_\lambda^2}.\ee
\end{proposition}

This is rather surprising, since the Weingarten functions of $O(N)$ do not involve the polynomials $\{N\}_\lambda$.

Propositions 3 and 4 imply that, both for $CO(N)$ and $CU(N)$ we have $\langle C_{ij}\rangle=\frac{\delta_{ij}}{N^2}$, so the average commutator is the identity matrix divided by $N^2$, as already known \cite{magee1,magee2,magee3}. Other examples of averages involving matrix elements are given in the text.

Let us mention that the quantity  $\frac{\chi_\lambda(\pi)}{d_\lambda}$, which appears prominently above, are called normalized characters and their asymptotics is considered as an important problem \cite{vershik,feray}. Moreover, when $N$ is large, both $F_N^U(\pi)$ and $F_N^O(\pi)$ are asymptotic to $N^{-2n}$ times $n!\sum_{\lambda\vdash n}\frac{\chi_\lambda(\pi)}{d_\lambda}$; this last quantity is the number of ways the permutation $\pi$ can be written as a commutator, in $S_n$.

\section{Unitary Group}

To avoid cumbersome notation, in this section $\langle\cdot\rangle$ denotes an average over the space of commutators in the unitary group, $CU(N)$.

\subsection{Powers of traces}

Let $\mathfrak{u}$ be the matrix representing $u$ in the irreducible representation labelled by some integer partition $\lambda$. The character of this representation is the Schur function 
\be \tr(\mathfrak{u})=s_\lambda(u)=\frac{\det(z_i^{N+\lambda_j-j})}{\det(z_i^{N-j})},\ee a symmetric polynomial of the $N$ eigenvalues $z_i$ of $u$ in the defining representation ($s_\lambda(u)=0$ unless $\ell(\lambda)\le N$). Therefore,
\be\label{schur1} \int_{U(N)} s_\lambda(uau^\dag b)du=\sum_{ijkl} \mathfrak{a}_{jk}\mathfrak{b}_{li}\int_{U(N)}\mathfrak{u}_{ij}\mathfrak{u}^*_{lk}du=\frac{s_\lambda(a)s_\lambda(b)}{s_\lambda(1_N)},\ee 
where we have used orthogonality of matrix elements $\int_{U(N)}\mathfrak{u}_{ij}\mathfrak{u}^*_{lk}du=\delta_{il}\delta_{jk}/s_\lambda(1_N)$, with $1_N$ being the identity matrix in dimension $N$. Moreover, we have the character orthogonality,
\be\label{schur2} \int_{U(N)} s_\lambda(u)s_\mu(u^\dag)du=\delta_{\lambda \mu}.\ee 

The above equations give the average value of the Schur function of the commutator as
\be\label{AS} \langle s_\lambda(C)\rangle=\frac{1}{s_\lambda(1_N)}.\ee On the other hand, the Weyl dimension formula gives $s_\lambda(1_N)=\frac{d_\lambda}{n!} [N]_\lambda$, with \cite{robin}
\be\label{Nl} [N]_\lambda=\prod_{i=1}^{\ell(\lambda)}\prod_{j=1}^{\lambda_i}(N+j-i),\ee where $d_\lambda$ is the dimension of the irreducible representation of the permutation group labelled by $\lambda$. Relation (\ref{AS}) allows the calculation of the average of any symmetric polynomial in the eigenvalues, because the Schur functions are a basis for that space. Power sum symmetric functions are written in terms of them as 
\be\label{p2s} p_\mu(C)=\sum_{\lambda\vdash n} \chi_\lambda(\mu)s_\lambda(C),\ee where $\chi_\lambda(\mu)$ are irreducible characters of the permutaion group. Using (\ref{AS}) we get our result.

As a particular case, we have \be \langle (\tr(C))^n\rangle =\sum_{\substack{\lambda\vdash n\\\ell(\lambda)\le N}}\frac{d_\lambda}{s_\lambda(1_N)}=n!\sum_{\substack{\lambda\vdash n\\\ell(\lambda)\le N}}\frac{1}{[N]_\lambda},\ee so that, for example,
\be \langle \tr(C)\rangle=\frac{1}{N},\ee
\be\langle (\tr(C))^2\rangle=\frac{4}{N^2-1},\ee
\be\langle (\tr(C))^3\rangle=\frac{18N}{(N^2-1)(N^2-4)}.\ee

Clearly $[N]_\lambda\sim N^n$ for large $N$ and fixed $n$. Hence we have the asymptotics
\be \langle (\tr(C))^n\rangle= \frac{n!p(n)}{N^n}+O(N^{-n-1}),\ee
where $p(n)$ is the number of partitions of $n$.

Since $\tr(C)$ is a complex number, it is also of interest to compute moments of its square modulus. However, that calculation is far more difficult. Using results for matrix elements in Proposition 3, we are able to find, for example, that
\be \left\langle \left|\tr(C)\right|^2\right\rangle=\frac{N^2}{N^2-1}.\ee
We can also compute the average value of $|\tr(C)|^4$, but it is too cumbersome to write here. The asymptotic series starts as
\be \left\langle \left|\tr(C)\right|^4\right\rangle= 2+\frac{4}{N^2}+\frac{2}{N^4}+O(N^{-6}).\ee These agree with the moments of a complex Gaussian distribution of mean $1/N$ and with variance given by $1/2$ for both the real and imaginary parts (notice that $\langle (\tr(C))^n\rangle$ vanishes like $N^{-n}$ for large $N$, while $\langle |\tr(C)|^{2n}\rangle$ remains finite).

\subsection{Traces of powers and eigenvalue densities}

Let the eigenvalues of $C$ be denoted $e^{i\theta_k}$ with $1\le k\le N$. 
Since the eigenphase density is even, it can be written as cosine Fourier series, \be\label{rcu}\rho_{CU(N)}(\theta)=\frac{1}{2\pi}+\sum_{n\ge 1}c_{N,n}\cos(n\theta).\ee The coefficients 
\be c_{N,n}=\frac{1}{\pi}\int_0^{2\pi} \cos(n\theta)\rho_{CU(N)}(\theta)d\theta\ee 
are related to $p_n(C)$ since 
\be \langle p_n(C)\rangle=\langle \tr(C^n)\rangle=\left\langle \sum_{k=1}^Ne^{in\theta_k}\right\rangle =N\pi c_{N,n}.\ee 

Using (\ref{p2s}) we have\be\label{trcn} \langle{\rm Tr}(C^n)\rangle=n!\sum_{\substack{\lambda\vdash n\\\ell(\lambda)\le N}}\frac{\chi_\lambda(n)}{d_\lambda [N]_\lambda}.\ee
The character $\chi_\lambda(n)$ is different from zero if and only if $\lambda$ is a hook partition, $\lambda=(n-k,1^k)$, in which case $\chi_\lambda(n)=(-1)^k$ and 
$d_\lambda={n-1\choose k}$. This gives
\be\langle{\rm Tr}(C^{n})\rangle=n\sum_{k=0}^{{\rm min}(N,n)-1}(-1)^kk!\frac{(n-k-1)!(N-k-1)!}{(N+n-k-1)!}, \ee
or 
\be\label{Q} \langle{\rm Tr}(C^{n})\rangle=\frac{n}{N}\sum_{k=0}^{{\rm min}(N,n)-1}(-1)^k{N-1 \choose k}^{-1}{N+n-k-1 \choose n-k-1}^{-1}. \ee
In particular,
\be \langle \tr(C^2)\rangle =-\frac{4}{N(N^2-1)},\ee
\be \langle \tr(C^3)\rangle =\frac{9(N^2+4)}{N(N^2-1)(N^2-4)},\ee
\be \langle \tr(C^4)\rangle =-\frac{64(4N^2+9)}{N(N^2-1)(N^2-4)(N^2-9)}.\ee

The $1/N$ series of the above quantities have integer coefficients, whose combinatorial interpretation has been investigated in \cite{magee2}.

The asymptotics of Eq.(\ref{Q}) for large $N$ is obtained by minimizing the binomials.  The leading term comes from the pair $(n=N,k=N-1)$, which gives
\be\label{rrr}\rho_{CU(N)}(\theta)=\frac{1}{2\pi}-\frac{(-1)^{N}}{N\pi}\cos(N\theta)+\frac{1}{N^2}\left(R_1+R_2+R_3\right),\ee with three kinds of remainder terms, $R_1=\sum_{n=1}^{N-1} c_{N,n}\cos(n\theta)$, $R_2=(c_{N,N}+\frac{(-1)^N}{N\pi})\cos(N\theta)$ and $R_3=\sum_{n=N+1}^{\infty} c_{N,n}\cos(n\theta)$, which can be bound as 
\be\label{rr1} |R_1|\le \sum_{n=1}^{N-1} \sum_{k=0}^{n-1} a_{N,n,k},\ee
\be |R_2|\le \sum_{k=0}^{N-2}a_{N,N,k}\ee
and
\be |R_3|\le \sum_{n=N+1}^{\infty} \sum_{k=0}^{N-1} a_{N,n,k},\ee
where $a_{N,n,k}=n{N-1 \choose k}^{-1}{N+n-k-1 \choose n-k-1}^{-1}$. What we now show is that all these remainder terms are $O(1)$, so that (\ref{rrr}) reduces to Eq.(\ref{arho1}).

Concerning $R_1$, we can see from inspection that $a_{N,1,0}=1$ and $a_{N,N-1,N-2}=O(1)$. These are the largest terms, all others in the sum (\ref{rr1}) are $O(N^{-1})$, because either $n{N-1 \choose k}=O(N)$ or ${N+n-k-1 \choose n-k-1}=O(N)$, or both. Since there are only $N-1$ terms in the sum, we have that $R_1=O(1)$.

Concerning $R_2$, we have $a_{N,N,k}\le a_{N,N,N-2}=\frac{N}{N^2-1}$, so $R_2\le N/(N+1)=O(1)$. This estimate can in fact be improved to $R_2=O(N^{-1})$, but we do not need that.

Concerning $R_3$, the sum over $n$ can be carried out to give $R_3=\sum_{k=0}^{N-1}b_{N,k}$, with $b_{N,k}=\frac{(2N-k)(N^2-k-1)}{(N-2)(N-1)}{N-1 \choose k}^{-1}{2N-k \choose N-k}^{-1}$. If $k=N-m$, then this is $O(N^{2-2m})$. Therefore, the term $m=1$ dominates the sum and $R_3=O(1)$.

\subsection{Matrix elements}

When dealing with invariant quantities, we could employ character theory. If we are interested in averages involving specific matrix elements of $C$, we must resort to some more algebra in terms of Weingarten functions, as indicated in (\ref{weingU}). For the unitary group, that function is
\be \Wg^{U}_N(\pi)=n!\sum_{\substack{\lambda\vdash n\\\ell(\lambda)\le N}}\frac{d_\lambda \chi_\lambda(\pi)}{[N]_\lambda}.\ee

Using that machinery, we can show that
\be \langle |C_{ii}|^2\rangle=\frac{N^2+N-1 }{(N^2-1)(N+1)}\ee
and that, whenever $i\neq j$,
\begin{align} 
\langle |C_{ij}|^2\rangle=\frac{N(N^2-2)}{(N^2-1)^2},\\
\langle C_{ik}C^*_{jk}\rangle=\frac{-1}{N(N^2-1)^2},\\
\langle C_{ii}C^*_{jj}\rangle=\frac{1}{(N^2-1)^2},
\end{align}
all other quadratic averages vanishing. 

For a random matrix $u\in U(N)$, it is known \cite{rmt} that the probability distribution of the modulus squared of any element, $z=|u_{ij}|^2$, is given by $(N-1)(1-z)^{N-2}$. Numerically, we find that this distribution is a very good approximation for elements of $C$, even for moderate $N$. The average value it predicts, $\int_0^1 \rho(z) zdz=\frac{1}{N}$, agrees with the correct expressions up to terms of order $N^{-4}$ for diagonal elements and up to terms of order $N^{-5}$ for non-diagonal elements.

The general correlation for matrix elements of the commutator (without any complex conjugation) is given by 
\be\label{XX}\left\langle \prod_{t=1}^n C_{i_t,j_t}\right\rangle=\sum_{\pi\in S_n}\delta_\pi(\vec{i},\vec{j})F_N^U(\pi),\ee where \be F_N^U(\pi)=n!\sum_{\substack{\lambda\vdash n\\\ell(\lambda)\le N}}\frac{\chi_\lambda(\pi)}{d_\lambda[N]_\lambda^2}.\ee

The simplest case is just
\be \langle C_{ij}\rangle =\frac{\delta_{ij}}{N^2},\ee which means that the average commutator in $U(N)$ is $\frac{1}{N^2}$ times the identity matrix. Slightly more complicated examples include
\be \langle C_{ij}C_{ji}\rangle=F_N^U((1)(2))\delta_{ij}+F_N^U((12)),\ee
\be \langle C_{ii}C_{jj}\rangle=F_N^U((1)(2))+F_N^U((12))\delta_{ij},\ee
\be \langle C_{ij}^2\rangle=(F_N^U((1)(2))+F_N^U((12)))\delta_{ij},\ee
where 
\be F_N^U((1)(2))=\frac{4}{N^2(N+1)^2}, \quad F_N^U((12)))=-\frac{8}{N(N^2-1)^2}.\ee

In order to prove Eq.(\ref{XX}), we start by writing $C_{\vec{i},\vec{j}}=\prod_{t=1}^n C_{i_t,j_t}$ in terms of the matrix elements of $u$ and $v$,
\be C_{\vec{i},\vec{j}}=\sum_{\vec{k}\vec{l}\vec{m}}\prod_t u_{i_tk_t}v_{k_tl_t}u^*_{m_tl_t}v^*_{j_tm_t},\ee whose average value is given by
\be \langle C_{\vec{i},\vec{j}}\rangle= \sum_{\vec{k}\vec{l}\vec{m}} \sum_{\sigma\tau\alpha\beta\in S_n}\Wg_N^U(\sigma^{-1}\tau)\Wg_N^U(\alpha^{-1}\beta)
\delta_{\sigma}(\vec{i},\vec{m})\delta_{\tau}(\vec{k},\vec{l})\delta_{\alpha}(\vec{k},\vec{j})\delta_{\beta}(\vec{l},\vec{m}).\ee 

Exchanging the order of the sums we can use identites like
\be\label{identi} \sum_{\vec{m}}\delta_{\sigma}(\vec{i},\vec{m})\delta_{\beta}(\vec{l},\vec{m})=\delta_{\sigma\beta^{-1}}(\vec{i},\vec{l})\ee
to get
\be \langle C_{\vec{i},\vec{j}}\rangle= \sum_{\sigma\tau\alpha\beta\in S_n}\Wg_N^U(\sigma^{-1}\tau)\Wg_N^U(\alpha^{-1}\beta)\delta_{\sigma\beta^{-1}\tau^{-1}\alpha}(\vec{i},\vec{j}).\ee
Changing variable from $\alpha$ to $\pi=\sigma\beta^{-1}\tau^{-1}\alpha$ we have
\be \langle C_{\vec{i},\vec{j}}\rangle= \sum_{\sigma\tau\pi\beta\in S_n}\Wg_N^U(\sigma^{-1}\tau)\Wg_N^U(\pi^{-1}\sigma\beta^{-1}\tau^{-1}\beta)\delta_{\pi}(\vec{i},\vec{j}).\ee
Writing the Weingarten functions in terms of characters and using (\ref{charac2}) we arrive at
\be\label{f} \langle C_{\vec{i},\vec{j}}\rangle= \frac{1}{n!}\sum_{\substack{\lambda\vdash n\\\ell(\lambda)\le N}}\frac{d_\lambda}{[N]_\lambda^2}\sum_{\tau\pi\beta\in S_n}\chi_\lambda(\tau\beta^{-1}\tau^{-1}\beta\pi^{-1})\delta_{\pi}(\vec{i},\vec{j}).\ee

It is interesting to notice the appearance of a permutation group commutator in this formula, $\tau\beta^{-1}\tau^{-1}\beta$. It is known (see exercise 7.68 in \cite{stanley}) that the number of ways to write some permutation $\alpha\in S_n$ as a commutator is given by 
\be\label{count} \#\{(\tau,\beta),\tau\beta^{-1}\tau^{-1}\beta=\alpha\}=n! \sum_{\eta\vdash n}\frac{\chi_\eta(\alpha)}{d_\eta}.\ee
Using this we can write 
\be\label{f2} \langle C_{\vec{i},\vec{j}}\rangle=\sum_{\pi\in S_n}\delta_{\pi}(\vec{i},\vec{j})\sum_{\substack{\lambda\vdash n\\\ell(\lambda)\le N}}\frac{d_\lambda}{[N]_\lambda^2}\sum_{\alpha\in S_n}\sum_{\eta\vdash n}\frac{\chi_\eta(\alpha)}{d_\eta}\chi_\lambda(\alpha\pi^{-1}),\ee
and the character orthogonality (\ref{charac2}) then gives the result
\be \langle C_{\vec{i},\vec{j}}\rangle=n!\sum_{\pi\in S_n}\delta_{\pi}(\vec{i},\vec{j})\sum_{\substack{\lambda\vdash n\\\ell(\lambda)\le N}}\frac{\chi_\lambda(\pi)}{d_\lambda[N]_\lambda^2}.\ee

\section{Orthogonal Group}

To avoid cumbersome notation, in this section $\langle\cdot\rangle$ denotes an average over the space of commutators in the orthogonal group, $CO(N)$.

\subsection{Powers of traces}

Matrices from $O(N)$ are real, so their eigenvalues are either real or come in complex conjugate pairs. For almost all $u\in O(N)$ the eigenvalues are complex, and one of them is equal to $1$ if $N$ is odd. For both $O(2M)$ and $O(2M+1)$, we denote the eigenvalues by $z_i$ and $\bar{z}_i$, with $1\le i\le M$.

Let $\tr(\mathfrak{u})=o_\lambda(u)$ denote the character of the orthogonal matrix $u$ in $O(2M)$ or $O(2M+1)$ in the irreducibe representation labelled by the partition $\lambda$ with $\ell(\lambda)< M$. For $O(2M)$, the irreducible characters are given by \cite{meckes}
\be o_\lambda(u)=\frac{\det(z_i^{M+\lambda_j-j}+\bar{z}_i^{M+\lambda_j-j})}{\det(z_i^{M-j}+\bar{z}_i^{M-j})}.\ee For $O(2M+1)$,
\be o_\lambda(u)=\frac{\det(z_i^{M+\lambda_j-j+1/2}-\bar{z}_i^{M+\lambda_j-j+1/2})}{\det(z_i^{M-j+1/2}-\bar{z}_i^{M-j+1/2})}.\ee
If $\ell(\lambda)=M$, the expression of the character must be multiplied by $2$.

In analogy with the unitary group, we have a simple expression for the average of the character computed for the commutator $C=uvu^Tv^T$:
\be\label{ao} \langle o_\lambda(C)\rangle=\frac{1}{o_\lambda(1_N)}.\ee
The Weyl dimension formula gives $o_\lambda(1_N)=\frac{d_\lambda}{n!}\{N\}_\lambda$ with \cite{samra}
\be\label{Nol} \{N\}_\lambda=\prod_{i=1}^{\ell(\lambda)}\prod_{j=1}^{{\rm min}(i,\lambda_i)}(N+\lambda_i+\lambda_j-i-j)\times \prod_{i=1}^{r}\prod_{j=i+1}^{\lambda_i}(N-\lambda'_i-\lambda'_j+i+j-2), \ee where $r$ is the size of the Durfee square of $\lambda$, i.e. the largest $i$ for which $\lambda_i-i\ge 0$. 

Besides these characters, it is convenient to define what are sometimes called `universal characters' \cite{murnaghan,newell,king,black,koike}. These are symmetric functions in the variables $\{z_1,z_1^{-1},...,z_M,z_M^{-1}\}$ labeled by partitions $\lambda$ for which the condition $\ell(\lambda)\leq M$ is not satisfied. They are equal to $o_{\widetilde{\lambda}},$ with a modified partition $\widetilde{\lambda}$. We shall make use of universal characters, with the understanding that anytime a function $o_\lambda$ is used in the context of a group $O(2M)$ or $O(2M+1)$ and $\ell(\lambda)>M$, it should be replaced by the modified version, $o_{\widetilde{\lambda}}$. We give some examples in Appendix B.

Power sum symmetric functions can be written in terms of $o_\lambda$. However, the expansion may involve polynomials of smaller degree. For $\mu\vdash n$,
\be\label{p2o} p_\mu(C)=\sum_{h=0}^{\left \lfloor{n/2}\right \rfloor} \sum_{\lambda\vdash n-2h}b_{\lambda}(\mu)o_\lambda(C).\ee The quantities $b_{\lambda}(\mu)$ are closely related to the theory of the Brauer algebra \cite{wenzl,hanlon}. When only actual characters are used to expand power sums, the coefficients have been called `Brauer characters'. Their theory was developed in more detail by Ram \cite{ram1,ram2}, where some families of explicit values were obtained. Since we use universal characters, we call $b_{\lambda}(\mu)$ the `universal Brauer characters'. The great advantage of using universal characters is that in this case $b_{\lambda}(\mu)$ does not depend on $N$ (see Appendix B).

Universal Brauer characters can be expressed in terms of Littlewood-Richardson coefficients \cite{koike}. If $\mu\vdash n$ and $\lambda\vdash n-2h$, then
\be b_{\lambda}(\mu)=\sum_{\nu \vdash n}\chi_\nu(\mu)\sum_{\beta\vdash h} c^{\nu}_{\lambda,2\beta},\ee where $2\beta$ is the partition whose parts are twice those of $\beta$. In particular, if $h=0$ and $\lambda\vdash n$, then $c^{\nu}_{\lambda,0}=\delta_{\nu,\lambda}$, so we have simply the permutation group characters:
\be b_{\lambda}(\mu)=\chi_\lambda(\mu) \text{ when } |\lambda|=|\mu|.\ee

As shown in \cite{sundaran}, for large $N$ the analogues of dimensions are:
\be b_\lambda(1^n)=\frac{n!}{(n-2h)!2^hh!}d_\lambda.\ee
Using this in (\ref{p2o}) leads to the explicit formula
\be \left\langle (\tr(C))^n\right\rangle=n!\sum_{h=0}^{\left \lfloor{n/2}\right \rfloor} \frac{1}{2^hh!}\sum_{\lambda\vdash n-2h}\frac{1}{\{N\}_\lambda}.\ee This gives, for example, 
\be \left\langle \tr(C)\right\rangle=\frac{1}{N},\ee
\be \left\langle (\tr(C))^2\right\rangle=\frac{N^3+N^2+2N+4}{(N-1)N(N+2)}=1+\frac{4}{N^2}+\frac{8}{N^4}+O(N^{-5}),\ee
\be \left\langle (\tr(C))^3\right\rangle=\frac{3N^4+9N^3-6N^2+18N+48}{(N-1)N(N^2-4)(N+4)}=\frac{3}{N}+\frac{18}{N^3}+O(N^{-5}).\ee 

The $1/N$ (or $1/(N-1)$) series of the above quantities have integer coefficients, whose combinatorial interpretation has been investigated in \cite{magee3}.

Since $\{N\}_\lambda\sim N^{|\lambda|}$ for large $N$ and fixed $n$, the dominating contribution to the first moments comes from the largest value of $h$. Therefore,
\be \left\langle (\tr(C))^{2n}\right\rangle= \frac{(2n)!}{2^nn!}+O(N^{-2})=(2n-1)!!+O(N^{-2})\ee and
\be \left\langle (\tr(C))^{2n-1}\right\rangle= \frac{(2n-1)!!}{N}+O(N^{-2}).\ee
This is consistent with $\tr(C)$ having distribution $\frac{1}{\sqrt{2\pi}}e^{-(x-\frac{1}{N})^2/2}$.

\subsection{Traces of powers and eigenvalue densities}

Let the eigenvalues of $C\in CO(2M)$ be denoted $e^{i\theta_k}$ and $e^{-i\theta_k}$ with $1\le k\le M$. The same holds for $C\in CO(2M+1)$, but in that case, an extra unit eigenvalue always exists. The non-zero eigenphases are described by a probability density, which as in the unitary case can be Fourier decomposed:
\be\rho_{CO(N)}(\theta)=\frac{1}{2\pi}+\sum_{n\ge 1}c_{N,n}\cos(n\theta).\ee The coefficients 
satisfy
\be\langle \tr(C^n)\rangle=\left\langle \sum_{k=1}^M \cos(n\theta_k)\right\rangle =N\pi c_{2M,n}\ee for $N=2M$ and  
\be\langle \tr(C^n)\rangle=\left\langle 1+\sum_{k=1}^M \cos(n\theta_k)\right\rangle =1+N\pi c_{2M+1,n}\ee for $N=2M+1$.

The first large $N$ correction to the constant density of states follow from the following asymptotic result.

\begin{proposition} 
If $N$ is large, then   
\be\label{Tless} \langle \tr(C^n)\rangle=
\begin{cases}1,\text{ if $n$ is even},\\
0, \text{ otherwise}\end{cases}+O(N^{-1})\ee if $N>n$ and 
\be \langle \tr(C^n)\rangle=
\begin{cases}0,\text{ if $N$ is even},\\
1, \text{ otherwise}\end{cases}+O(N^{-1})
\ee if $N\le n$.
\end{proposition}

Proposition 2 is a direct consequence of Proposition 5, since
\begin{align} \rho_{CO(N)}(\theta)&=\frac{1}{2\pi}+\frac{1}{\pi N}\sum_{n=1}^{M-1}\cos(2n\theta)+O(N^{-2})
\\&=\frac{1}{2\pi}-\frac{1}{2\pi N}+\frac{\sin((N-1)\theta)}{2\pi N\sin(\theta)}+O(N^{-2})\end{align} for $N=2M$ and 
\begin{align} \rho_{CO(N)}(\theta)&=\frac{1}{2\pi}-\frac{1}{\pi N}\sum_{n=0}^{M-1}\cos((2n+1)\theta)+O(N^{-2})
\\&=\frac{1}{2\pi}-\frac{\sin((N-1)\theta)}{2\pi N\sin(\theta)}+O(N^{-2})\end{align} for $N=2M+1$.

Taking the average value of equation (\ref{p2o}) in this case leads to
\be\label{pntol}\langle{\rm Tr}(C^{n})\rangle=\sum_{h=0}^{\left \lfloor{n/2}\right \rfloor} \sum_{\lambda\vdash n-2h}\frac{b_{\lambda}(n)}{o_\lambda(1^N)},\ee 
but this can in fact be much simplified, because of 
\begin{proposition} 
For all $\lambda\vdash n-2h$ with $0<h<n/2$, the universal Brauer charater vanishes, $b_{\lambda}(n)=0$.
\end{proposition}

In order to prove this, we start from 
\be b_{\lambda}(n)=\sum_{\nu \vdash n}\chi_\nu(n)\sum_{\beta\vdash h} c^{\nu}_{\lambda,2\beta}.\ee The character $\chi_\nu(n)$ is different from zero if and only if $\nu$ is a hook partition, in which case $\chi_\nu(n)=(-1)^{v_1}$, where $v_1$ is the number of $1$'s in $\nu$. On the other hand, it is known that, if $\nu$ is a hook, the quantity $c^\nu_{\lambda,2\beta}$ is different from zero only if both $\lambda$ and $2\beta$ are also hooks. But $2\beta$ can only be a hook if $\beta=(h)$. Therefore,
\be\label{pieri} b_{\lambda}(n)=\sum_{\substack{\nu \vdash n \\ \text{hook}}}(-1)^{v_1}c^{\nu}_{\lambda,(2h)}.\ee

By the Pieri rule, $c^\nu_{\lambda,(2h)}=1$ if and only if the Young diagram of $\nu$ can be obtained from the Young diagram of $\lambda$ by adding $2h$ boxes to it, no two in the same column. Since $\nu$ is a hook, there only two possibilities: 1) all boxes are added to the first line; 2) one box is added as a new line. These possibilities cancel when computing (\ref{pieri}).

Therefore, \be\label{trn}\langle{\rm Tr}(C^{n})\rangle= \sum_{\lambda\vdash n}\frac{\chi_{\lambda}(n)}{o_\lambda(1^N)}
+\begin{cases} 1,\text{ if $n$ even},\\0,\text{ if $n$ odd},\end{cases}\ee the last term coming from $b_{\emptyset}$, one of the families of Brauer characters that have been computed by Ram.

For example,
\be \left\langle \tr(C^2)\right\rangle=\frac{N^3+N^2-2N-4}{N(N-1)(N+2)}=1-\frac{4}{N^3}+O(N^{-4}),\ee
\be \left\langle \tr(C^3)\right\rangle=\frac{9N^2+27N+36}{N(N-2)(N-1)(N+2)(N+4)}=\frac{9}{N^3}+O(N^{-4}).\ee

We now prove Proposition 5. As in the unitary case, we use that $\chi_{\lambda}(n)$ is different from zero if and only if $\lambda$ is a hook partition, $\lambda=(n-k,1^k)$, in which case $\chi_\lambda(n)=(-1)^k$ and $d_\lambda={n-1\choose k}$. In that case we have
\be \{N\}_{(n-k,1^k)}=\frac{(N+2n-2k-2)(N+n-k-2)!}{(N+n-2k-2)(N-k-2)!}.\ee

\subsubsection{Case $N>n$}

Comparing (\ref{trn}) with (\ref{Tless}), we just have to show that $Q(N,n)=\sum_{\lambda\vdash n}\chi_{\lambda}(n)/o_\lambda(1^N)$ is $O(N^{-1})$. The explicit form of $Q(N,n)$ is
\be n\sum_{k=0}^{n-1}(-1)^k{N-1 \choose k}^{-1}{N+n-k-2 \choose n-k-1}^{-1}\frac{N+n-2k-2}{(N-k-1)(N+2n-2k-2)}. \ee

The binomial in the middle shows that the largest term in the sum is always the last one, $k=n-1$, and it is equal to ${N \choose n}^{-1}$. We distinguish three cases: i) $n=1$; then $Q(N,1)=N^{-1}$. ii) $1<n<N-1$; the largest term in the sum is $O(N^{-2})$; remaining ones are $O(N^{-3})$; there are less than $N$ terms in the sum, so $Q(N,n)=O(N^{-2})$. iii) $n=N-1$; largest term is $N^{-1}$, second largest ($k=n-2=N-3$) is $O(N^{-3})$; therefore, $Q(N,N-1)=O(N^{-1})$. 

We must notice that when $\ell(\lambda)>M$ and $N=2M$ or $2M+1$, the function $o_\lambda$ must be replaced by the modified version, $o_{\widetilde{\lambda}}$. However, the modification of a hook is still a hook (see Appendix B), and the above analysis still holds, with straighforward adjustments.

\subsubsection{Case $N\le n$}

The major difference with the situation $N>n$ is that now it may happen that $\widetilde{\lambda}=\emptyset$, in which case there may be a finite contribution to $Q(N,n)$ for large $N$, because $o_\emptyset=1$.

It is only possible to have $\widetilde{\lambda}=\emptyset$ when $2\ell(\lambda)-N=n$, and this can only be satisfied when $N$ and $n$ are either both even or both odd. Then $o_\lambda=(-1)^{n-\ell(\lambda)}o_{\widetilde{\lambda}}$ and the value of $\chi_\lambda(n)$ is $(-1)^{\ell(\lambda)-1}$, so the total contribution is $(-1)^{n-1}$.

In the end, we have that, for large $N$, the value of $Q(N,n)$ 
approaches
\be \begin{cases} -1,&\text{ if $n$ is even},\\0, &\text{ otherwise} \end{cases},\text{ if $N$ is even}\ee and \be \begin{cases} 0,&\text{ if $n$ is even},\\1, &\text{ otherwise} \end{cases}, \text{ if $N$ is odd}.\ee

Combining this with (\ref{trn}) leads to (\ref{Tless}).

As we have seen in Section 4.2.1, the partitions $\lambda$ which give the dominating contribution to $Q(N,n)$ are always the longest ones. The length of $\lambda$ was $k+1$ in Section 4.2.1, and this could be as large as $n$. When $n\ge N$ and $N=2M$ or $2M+1$, the length of non-modified partitions $\lambda$ is restricted to be $\ell(\lambda)\le M$, so the corrections to $Q(N,n)$ coming from them are even smaller than they were for $N>n$. 

On the other hand, modified partitions are of the form $\widetilde{\lambda}=(n-k,1^{N-k-2})$, with $M< k< N-1$. Then $d_{\widetilde{\lambda}}={N+n-2k-3\choose N-k-2}$ and the contribution from this partition to $Q(N,n)$ is
\be \frac{(N+n-2k-2)}{(N+2n-2k-2)}\frac{n}{(N-k-1)}{N+n-k-2\choose n-k-1}^{-1}{N-1\choose k}^{-1}.\ee This is $O(N^{-2})$ for $(n=N,k=N-2)$ and of higher order in $1/N$ for other values. Therefore, corrections to the leading results of Proposition 2 are $O(N^{-2})$.

\subsection{Matrix elements} 

Using Weingarten functions, it is straightforward to show 
\be \langle C_{ij}\rangle=\frac{\delta_{ij}}{N^2},\ee so the average commutator for $O(N)$ is also $\frac{1}{N^2}$ times the identity matrix.

The general average $\langle C_{\vec{i},\vec{j}}\rangle$ is rather hard to treat. We could find an explicit form only in the particular case when the strings $\vec{i}$ and $\vec{j}$ do not contain repeated elements and are thus related by a single permutation. In that case we have
\be\label{CO2} \langle C_{\vec{i},\pi(\vec{i})}\rangle=F_N^O(\pi)=n!\sum_{\substack{\lambda\vdash n\\\ell(\lambda)\le N}}\frac{\chi_\lambda(\pi)}{d_\lambda\{N\}_\lambda^2}.\ee 

This gives, for example,
\be \left\langle C_{11}C_{22}\right\rangle=F_N^O((1)(2))=\frac{4(N^2+2N+2)}{(N-1)^2N^2(N+2)^2},\ee
and
\be \left\langle C_{12}C_{21}\right\rangle=F_N^O((12))=-\frac{8(N+1)}{(N-1)^2N^2(N+2)^2}.\ee

The distribution of the matrix elements of orthogonal matrices $z=u_{ij}$ is known \cite{rmt} to be proportional to $(1-z^2)^{(N-3)/2}$. Numerically we find that this is a very good approximation for elements of the commutator, however we are unable to compute moments higher than the first because of the restriction that the indices must not be repeated.

We now prove Eq.(\ref{CO2}).

We start by writing
\be  C_{\vec{i},\vec{j}}=\sum_{\vec{k},\vec{l},\vec{m}}u_{\vec{i}\vec{k}}v_{\vec{k}\vec{l}}u_{\vec{m}\vec{l}}v_{\vec{j}\vec{m}},\ee and using
\be \left \langle u_{\vec{i}\vec{k}}u_{\vec{m}\vec{l}}\right\rangle =\sum_{\sigma,\tau\in M_n}
\Delta_\sigma(\vec{i}\diamond\vec{m})\Delta_{\tau^{-1}}(\vec{l}\diamond\vec{k})\Wg^{O}_N(\sigma^{-1}\tau)\ee
and 
\be \left \langle v_{\vec{k}\vec{l}}v_{\vec{j}\vec{m}}\right\rangle =\sum_{\alpha,\beta\in M_n}
\Delta_{\beta^{-1}}(\vec{m}\diamond\vec{l})\Delta_\alpha(\vec{k}\diamond\vec{j})\Wg^{O}_N(\alpha^{-1}\beta),\ee where the arguments of the $\Delta$ functions are interleaving operations,
\be \vec{i}\diamond\vec{m}=(i_1,m_1,i_2,m_2,...,i_n,m_n).\ee

The quantity $\sum_{\vec{m}}\Delta_\sigma(\vec{i}\diamond\vec{m})\Delta_{\beta}(\vec{m}\diamond\vec{l})$ is closely related to the Brauer product of the matchings $\sigma$ and $\beta$ (see Appendix). It is given by
\be \sum_{\vec{m}}\Delta_\sigma(\vec{i}\diamond \vec{m})\Delta_{\beta}(\vec{m}\diamond\vec{l})=N^{{\rm loops}(\sigma,\beta)}
\Delta_{\sigma\circ \beta}(\vec{i}\diamond\vec{l}),\ee where $\sigma\circ\beta$ is the Brauer product and ${\rm loops}(\sigma,\beta)$ is the number of loops produced when the diagrams of $\sigma$ and $\beta$ are joined. The multiple sum over lists $\vec{k},\vec{l},\vec{m}$ will therefore lead to something proportional to $\Delta_\pi(\vec{i}\diamond\vec{j})$ with $\pi=\sigma\circ\beta^{-1}\circ\tau^{-1}\circ\alpha$. 

We found this problem to be tractable only under the assumption that $\vec{i}$ and $\vec{j}$ do not contain repeated elements. This means that the diagram of the matching $\pi$ only has vertical lines and can be represented by a permutation in $S_n$. This in turn implies that the diagrams of $\sigma,\beta,\tau,\alpha$ all consist exclusively of vertical lines and can all be represented by permutations in $S_n$ (they are permutational). In this case the number of loops is zero for all the joinings. 

From now on we understand that $\pi,\sigma,\beta,\tau,\alpha\in S_n$ and $\Delta_\pi(\vec{i}\diamond\vec{j})=\delta_\pi(\vec{i},\vec{j})$. Notice that we can replace the Brauer product by the usual permutation product, e.g. $\sigma\circ\beta^{-1}=\sigma\beta^{-1}$. Moreover, the coset type coincides with the cycle type (see Appendix).

Solving $\pi=\sigma\beta^{-1}\tau^{-1}\alpha$ for $\sigma$ we have 
\be\langle C_{\vec{i},\pi(\vec{i})}\rangle=\sum_{\tau,\alpha,\beta\in S_n}\Wg^{O}_N(\beta^{-1}\tau^{-1}\alpha\pi^{-1}\tau)\Wg^{O}_N(\alpha^{-1}\beta),\ee or, after some relabelling,
\be \langle C_{\vec{i},\pi(\vec{i})}\rangle=\sum_{\tau,\alpha,\beta\in S_n}\Wg^{O}_N(\tau^{-1}\beta\tau\beta^{-1}\alpha\pi^{-1})\Wg^{O}_N(\alpha^{-1}).
\ee
A permutation group commutator has appeared again. Using (\ref{count}) we have
\be\label{inter} \langle C_{\vec{i},\pi(\vec{i})}\rangle=\sum_{\rho,\alpha\in S_n}\Wg^{O}_N(\rho\alpha\pi^{-1})\Wg^{O}_N(\alpha^{-1})n!\sum_{\lambda\vdash n}\frac{\chi_\lambda(\rho)}{d_\lambda}.\ee

A result conjectured in \cite{lucas} and later proved in \cite{chapuy} implies that
\be \sum_{\rho\in S_n}\Wg^{O}_N(\rho\alpha\pi^{-1})\chi_\lambda(\rho)=\frac{\chi_\lambda(\alpha\pi^{-1})}{\{N\}_\lambda}.\ee This can be used twice to give  indeed
\be \langle C_{\vec{i},\pi(\vec{i})}\rangle=n!\sum_{\substack{\lambda\vdash n\\\ell(\lambda)\le N}}\frac{\chi_\lambda(\pi)}{d_\lambda\{N\}_\lambda^2}.\ee 

\section*{Acknowledgments} This research was supported by grants 400906/2016-3 and 306765/2018-7 from CNPq. P.H.S.P. and M.R.B were supported by fellowships from Capes and Fapemig. We thank Prof. Ronald King for his help with understanding the universal characters of the orthogonal group. We were fortunate to have a very attentive anonymous reviwer, whose input helped improve the paper. Now new data was generated during this study.

\appendix

\renewcommand{\theequation}{\thesection\arabic{equation}}

\section{The Brauer monoid}
\setcounter{equation}{0}

Let $S_n$ be the group of all permutations acting on the set $[n]:=\{1,...,n\}$ and let $\mathcal{C}_\lambda$ denote the conjugacy class of all permutations with cycle type $\lambda$.
Irreducible representations of $S_n$ are also labelled by such partitions, and $\chi_\lambda(\mu)$ is the character of the cycle type $\mu$ in the irrep $\lambda$, with $\chi_\lambda(1)$ being the dimension of the irrep.

Characters satisfy two orthogonality relations, 
\be\label{charac1} \sum_{\mu\vdash
n}\chi_\mu(\lambda)\chi_\mu(\omega)=\frac{n!}{|\mathcal{C}_\lambda|}\delta_{\lambda,\omega}, \quad
\frac{1}{n!}\sum_{\lambda\vdash n}|\mathcal{C}_\lambda| \chi_\mu(\lambda) \chi_\omega(\lambda)=
\delta_{\mu,\omega}.\ee The latter may be generalized as a sum over permutations as \be\label{charac2}
\frac{1}{n!}\sum_{\pi\in S_n}\chi_\mu(\pi) \chi_\lambda(\pi\sigma)=
\frac{\chi_\lambda(\sigma)}{\chi_\lambda(1)}\delta_{\mu,\lambda}.\ee
 
A matching on the set $[2n]$ is a collection of $n$ disjoint subsets with two elements each (`blocks'), such as \be\mathfrak{t}:=\{\{1,2\},\{3,4\},...,\{2n-1,2n\}\}.\ee The above matching is the `trivial' one. The set of all matchings on $[2n]$ is $M_n$, and $\pi\in S_{2n}$ acts on $M_n$ by simply replacing $i$ by $\pi(i)$. The hyperoctahedral group $H_n\subset S_{2n}$ is $H_n=\{h\in S_{2n}, h(\mathfrak{t})=\mathfrak{t}\}$. The elements of the coset $S_{2n}/H_n$ are in bijection with matchings: to $\mathfrak{m}$ we associate $\sigma$ that satisfies $\sigma(\mathfrak{t})=\mathfrak{m}$ and also $\sigma(2i-1)<\sigma(2i)$ and $\sigma(1)<\sigma(3)<\cdots\sigma(2n-1)$ (see \cite{matsumoto}). 

\begin{figure}[t!]
\center
\includegraphics[scale=1,clip]{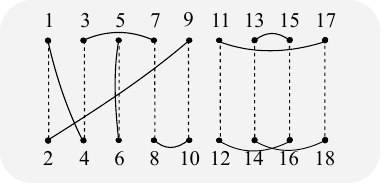}\hspace{1cm}\includegraphics[scale=0.4,clip]{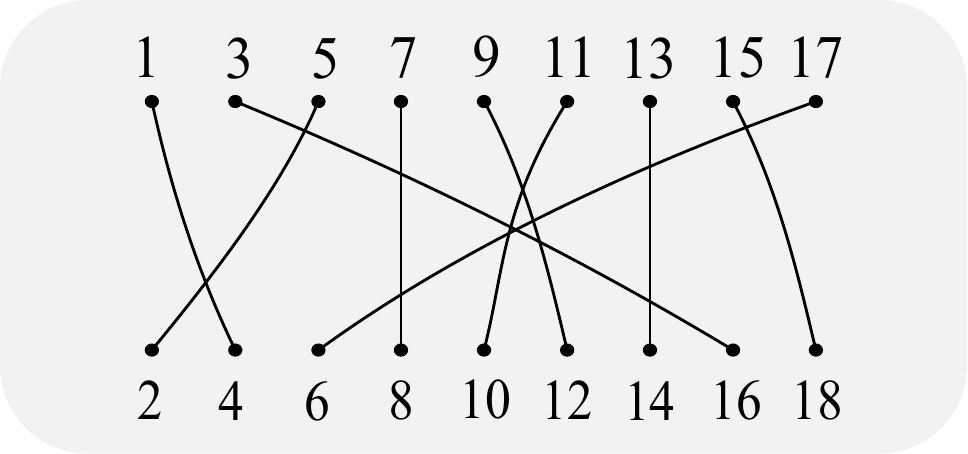}
\caption{Left: Diagrammatic representation of the matching $\sigma$ given in the text. Right: Diagrammatic representation of the permutational matching $\tau$ given in the text.} 
\end{figure}

Given $\sigma\in S_{2n}/H_n$ and $\vec{i}=(i_1,\ldots,i_{2n})$, define the function \be\label{Delta} \Delta_\sigma(i)=\prod_{k=1}^{n} \delta_{i_{\sigma(2k-1)},i_{\sigma(2k)}},\ee which is equal to 1 if and only if the elements of the sequence $i$ are pairwise equal according to the matching associated with $\sigma$. For example, $(1,1,2,2,...,n,n)$ satisfies the trivial matching.

Matchings may be represented by diagrams such as the one in Figure 2. The dashed lines indicate how the points are connected according to the trivial matching, while the solid lines indicate how they are connected according to the matching $\sigma=\{\{1,4\},\{2,9\},\{3,7\},\{5,6\},\{8,10\},\{11,17\},\{12,16\},\{13,15\},\{14,18\}\}.$ The coset type of a matching is the list of half the sizes of connected components in the diagram. It is $(4,4,1)$ for this $\sigma$. The Weingarten function 
$\Wg^{O}_N(\sigma)$ depends only on the coset type of $\sigma$.

Notice that if in a certain matching odd numbers are matched only to even numbers, so every block is of the form $\{2i-1,2j\}$, then this matching can be associated with the permutation taking $i$ to $j$ and the corresponding diagram has only vertical lines. Let us say that these matchings are permutational. For example, the matching $\tau=\{\{1,4\},\{2,5\},\{3,16\},\{6,17\},\{7,8\},\{9,12\},\{10,11\},\{13,14\},\{15,18\}\}$ is permutational; the permutation associated with it is $(1\,2\,8\,9\,3)(4)(5\,6)(7)$. The coset type and cycle type are clearly equal for permutational matchings; both are given by $(5,2,1,1)$ for this $\tau$.

\begin{figure}[t!]
\center
\includegraphics[scale=0.7,clip]{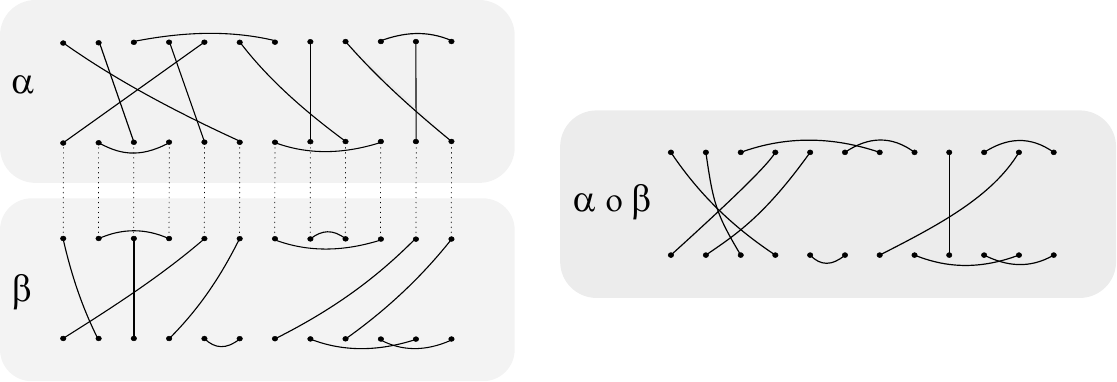}\hspace{1cm}
\caption{The Brauer product $\alpha\circ \beta$ is shown on the right for the matchings $\alpha$ and $\beta$ on the left. Loops are ignored.} 
\end{figure}

A product may be defined for matchings by simply joining the diagrams, as in Figure 4. In order to multiply $\alpha$ by $\beta$ from the right, the top points of $\beta$ are identified with the bottom points of $\alpha$. In this process some loops are created involving such points, two in this example. These loops are ignored in the product, $\alpha \circ \beta$, which is also a matching. We denote by loops$(\alpha,\beta)$ the number of loops created. (With this product $\circ$, matchings form a monoid; the Brauer algebra, on the other hand, involves a different product, $\alpha \times \beta=N^{{\rm loops}(\alpha,\beta)}\alpha\circ\beta$, which keeps track of the loops. We do not use this structure.) 

Clearly, if $\alpha$ and $\beta$ are both permutational matchings, than the matching $\alpha\circ\beta$ is also permutational, and the corresponding permutation is given by the usual permutation product $\alpha\beta$.

\section{Universal characters of $O(N)$}
\setcounter{equation}{0}

Functions $o_\lambda$ correspond to irreducible characters of $O(2M)$ or $O(2M+1)$ only when $\ell(\lambda)=\lambda'_1\le M$. The set of such actual characters forms a basis for the space of symmetric functions on variables $\{z_1,z_1^{-1},\dotsc,z_M,z_M^{-1}\}$. Unfortunately, the coefficients in the expansion of power sums in terms of them will in general depend on $M$. 

For example, let us consider the problem of expanding $p_4$ in terms of the characters of $O(2M)$. It turns out that 
\be\label{p42} p_4=o_4 \text{ for }O(2),\ee  
\be\label{p44} p_4=o_4-o_{31}+o_2\text{ for } O(4),\ee
\be\label{p46} p_4=o_4-o_{31}+o_{211}-o_{11}+o_\emptyset\text{ for }O(6)\ee and 
\be\label{p48} p_4=o_4-o_{31}+o_{211}-o_{1111}+o_\emptyset\text{ for $O(2M)$ with $M\ge 4$}.\ee 
Notice how $\ell(\lambda)\le M$ for all terms in these expansions.

A partition $\lambda$ with $\ell(\lambda)>M$ does not define an irreducible representation of $O(2M)$ or $O(2M+1)$, so we should not associate with it an actual character. However, it can still be associated to a symmetric function, $o_{\widetilde{\lambda}}$, by means of a modified partition defined as follows \cite{king,black}. For $O(N)$, let $m=2\ell(\lambda)-N$.  Then remove from the Young diagram of $\lambda$ a total of $m$ adjacent boxes, starting from the bottom of the first column and keeping always at the boundary of the diagram. In this way the changes to be implemented are, in sequence, $\lambda'_1\to\lambda'_2-1$, then $\lambda'_2\to\lambda'_3-1$, etc. until the procedures stops at some column $c$. If $m$ is too large and there are not enough boxes to accommodate this procedure, or if the remaining diagram is not a partition, then $o_\lambda=0$; otherwise $o_\lambda=(-1)^{c-1}o_{\widetilde{\lambda}}$.
 
\begin{figure}[t]
\center
\includegraphics[scale=0.7,clip]{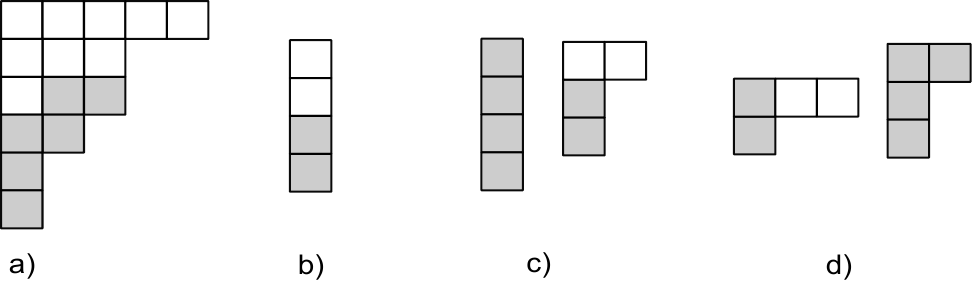}\hspace{1cm}
\caption{Modification procedure. In a) we have $\lambda=(5,3,3,2,1,1)$ and $O(6)$ so we remove $12-6=6$ adjacent boxes, ending at column $c=3$, and we are left with $\widetilde{\lambda}=(5,3,1)$. Hence $o_{(5,3,3,2,1,1)}=o_{(5,3,1)}$ for $O(6)$. In b)-d) we have the modifications required to expand $p_4$ in terms of universal characters, as explained in the text.} 
\end{figure}

In Figure 4 we show some examples. A generic case appears in Figure 4.a. When $\lambda=(1,1,1,1)$ and $N=6$, we must remove $8-6=2$ boxes as in Figure 4.b, in which case we get $c=1$ and $\widetilde{\lambda}=(1,1)$. Hence, for $O(6)$ we have $o_{1,1,1,1}=o_{1,1}$ and the universal relation (\ref{p48}) reduces to (\ref{p46}). When $N=4$, we must remove $8-4=4$ boxes from $\lambda=(1,1,1,1)$, as in Figure 4.c. We end up with no boxes at all, so $o_{1,1,1,1}=o_\emptyset$ for $O(4)$. Figure 4.c also shows how  we must remove $6-4=2$ boxes from $\lambda=(2,1,1)$, arriving at $o_{2,1,1}=o_2$. We see that (\ref{p48}) indeed reduces to (\ref{p44}). Finally, take $N=2$. We cannot remove $8-2=6$ boxes from $(1,1,1,1)$, so $o_{1,1,1,1}=0$ for this group; when we remove $4-2=2$ boxes from $(3,1)$, the result (Figure 4.d) is not the diagram of a partition, so $o_{3,1}=0$ for this group; removing $6-2=4$ boxes from $(2,1,1)$ leads to the empty partition, and the removing procedure ends in the second column so $c=2$, hence $o_{2,1,1}=-o_\emptyset$ for this group. Thereby (\ref{p48}) reduces to (\ref{p42}). 

We must remark that the modification rule to get $\widetilde{\lambda}$ from $\lambda$ is incorrectly stated in \cite{meckes} and \cite{ram1}.

With the appropriate use of modifications, the expansion (\ref{p48}) is valid for all values of $N$. In fact, the coefficients $b_{\lambda}(\mu)$ in the expansion of power sums are then independent of $N$ \cite{koike,kingf}. In particular, in the expansion of a power sum $p_n$ only hook partitions are involved. For $O(N)$, the modification of $\lambda=(n-k,1^k)$, when $\ell(\lambda)=k+1$, will require the removal of $m=2k+2-N$ boxes. This will lead to $(n-k, 1^{k-m})=(n-k, 1^{N-k-2})$, if $m\le k$, or to the empty partition if $m=n$.

Let us mention that, when a Schur function $s_\mu$ is decomposed in terms of $o_\lambda$, which corresponds to the branching rule of $U(N)\supset O(N)$, it happens that $\ell(\lambda)\le \ell(\mu)$. If $N=2M$ or $N=2M+1$ and $\ell(\mu)\le M$, only actual characters appear in the branching rule. This is no longer true if $\ell(\mu)>M$, and then universal characters must also be used if the coefficients are to be independent of $N$ (see \cite{koike,branch}). Curiously, this situation is not discussed in the classic book \cite{fulton}, for example, where only the case $\ell(\mu)\le M$ appears.

\end{document}